\documentclass[letterpaper,12pt]{article}
\usepackage{color,amssymb,enumerate,float}
\usepackage{amsmath}
\usepackage{cancel}

\usepackage{ifpdf}
\ifpdf
	\usepackage[pdftex]{graphicx}
	\usepackage[pdftex,unicode,implicit]{hyperref}

	\hypersetup{
  	pdftitle     = {}, 
  	pdfkeywords  = {},
  	pdfauthor    = {},
  	pdfcreator   = {pdf\LaTeXe\ with package \flqq hyperref\frqq},
  	pdfproducer  = {pdf\LaTeXe\ with package \flqq hyperref\frqq},
  	pdfpagemode  = UseNone,  
  	pdffitwindow = true,  
  	unicode      = true,
  	plainpages   = true,
  	colorlinks   = true,  
  	citecolor    = black,  
  	urlcolor     = blue, 
  	linkcolor    = black
	}

\else

  \usepackage[dvips]{graphicx}
  
\fi 

\makeatletter
\@addtoreset{equation}{section}
\makeatother

\begin{document}
	
\thispagestyle{empty}

\begin{center}
	{\bf \LARGE Renormalization of the nonprojectable Ho\v{r}ava theory}
	\vspace*{15mm}
	
	{\large Jorge Bellor\'{\i}n}$^{1,a}$,
	{\large Claudio B\'orquez$^{2,b}$}
	{\large and Byron Droguett}$^{1,c}$
	\vspace{3ex}
	
	$^1${\it Department of Physics, Universidad de Antofagasta, 1240000 Antofagasta, Chile.}
	
	$^2${\it Facultad de Ingenier\'ia, Arquitectura y Diseño, Universidad San Sebasti\'an, Lago Panguipulli 1390, Puerto Montt, Chile.}
	
	\vspace*{2ex}
	$^a${\tt jorge.bellorin@uantof.cl,} \hspace{1em}
	$^b${\tt claudio.borquez@uss.cl,} \hspace{1em}
	$^c${\tt byron.droguett@uantof.cl}

	\vspace*{15mm}
	{\bf Abstract}
\begin{quotation}{\small}
		We present the proof of renormalization of the Ho\v{r}ava theory, in the nonprojectable version. We obtain a form of the quantum action that exhibits a manifest BRST-symmetry structure. Previous analysis has shown that the divergences produced by irregular loops cancel completely between them. The remaining divergences are local. The renormalization is achieved by using the approach developed by Barvinsky et al.~with the background-field formalism. 
\end{quotation}
\end{center}

\newpage
\section{Introduction}
In this paper we present the proof of renormalization of the Ho\v{r}ava theory, considering its nonprojectable version \cite{Horava:2009uw,Blas:2009qj}. This theory, whose gauge group is given by the foliation-preserving diffeomorphisms (FDiff), is a proposal for a quantum theory of gravitation. The theory is unitary \cite{Bellorin:2022efu}, we are presenting here its renormalization, and it might yield the classical dynamics of general relativity at the large-distance limit. 

We base the proof on three main aspects. First, the theory is quantized under the Batalin-Fradkin-Vilkovisky (BFV) formalism, incorporating the second-class constraints \cite{Fradkin:1975cq,Batalin:1977pb,Fradkin:1977xi}. This formalism allows us to introduce a local gauge-fixing condition that leads to regular propagators for most of the fields \cite{Barvinsky:2015kil,Bellorin:2021udn}, together with the measure of the second-class constraints \cite{Senjanovic:1976br,Fradkin1973}. After the integration on some ghost fields and the redefinition of the Becchi-Rouet-Stora-Tyutin (BRST) symmetry transformations, we get a form of the quantum action with manifest BRST-symmetry structure. Second, it is known from previous studies \cite{Bellorin:2022qeu,Bellorin:2023dwk} that the only divergences produced by the integration along the frequency (called irregular loops) cancel exactly between them. In the integration on the spatial momentum, the behavior of the irregular propagators is equivalent to the regular ones. Hence, all divergences are local \cite{Anselmi:2007ri,Anselmi:2008bq}. The highest superficial degree of divergence is equal to the order of the bare Lagrangian. Third, we use the approach developed by Barvinsky et al.~\cite{Barvinsky:2017zlx} to undertake the renormalization of gauge theories, which is based on the background-field formalism \cite{DeWitt:1967ub,Abbott:1981ke}.

The proof of renormalization of the projectable case, which is a version of the Ho\v{r}ava theory defined by the restriction that the lapse function depends exclusively on time, is known \cite{Barvinsky:2015kil,Barvinsky:2017zlx}. The need for an  anisotropic gauge-fixing condition that leads to regular propagators was identified in this case. The resulting Lagrangian is local when it is expressed in terms of canonical variables. However, a fundamental difference in the quantization of the projectable and nonprojectable cases is the absence of second-class constraints in the former. In the nonprojectable case, a similar quantization can be carried on, with the analogous gauge-fixing condition. The second-class constraints lead to a modification of the measure of the path integral. The measure has the effect of yielding irregular propagators on some auxiliary fields, despite the fact that the rest of the quantum fields acquire regular propagators. Since the regular structure is important for the control of divergences \cite{Anselmi:2008bq}, a careful study of the consequences of the irregular propagators is required. For this reason it becomes essential for the previously mentioned analysis to show not only the divergences produced by the irregular loops are canceled, but also the fact that these are the only divergences in the direction of the frequency.

As most of the modern approaches of renormalization of gauge theories, the proof relies on the BRST symmetry and the background-field formalism. The Slavnov-Taylor and Ward identities are useful to determine the divergences of the effective action. On the other hand, the BFV quantization is based on the Hamiltonian formalism. Therefore, it is important to arrive at a form of the quantum action being separated in the usual way: a sector invariant under the FDiff gauge symmetry, and another sector fixing the gauge symmetry by means of the BRST operator. We find such a BRST-symmetry structure. This allows us to apply the background-field formalism of Ref.~\cite{Barvinsky:2017zlx}.

The analysis of renormalization requires one to know the propagators. With the vertices, the feature that we require is the highest order in spatial derivatives, independent	 of their explicit form. For this reason, in this study we need only to write explicitly the higher-order terms in the Lagrangian that contribute to the propagators.


\section{BFV Quantization of the Ho\v{r}ava theory}
\subsection{The classical theory}
The initial assumption in the definition of he Ho\v{r}ava theory is the existence of a foliation of spatial slices along a given direction of time with absolute meaning. In the classical theory, the fields representing the gravitational interaction are the Arnowitt-Deser-Misner (ADM) field variables $N(t,\vec{x})$, $N^i(t,\vec{x})$, and $g_{ij}(t,\vec{x})$. We deal only with the nonprojectable version, on which the lapse function $N$ can depend on time and space. The corresponding gauge symmetry is given by the FDiff. In terms of a given coordinate system $(t,\vec{x})$, the FDiff acts infinitesimally as $\delta t = f(t)$ and $\delta x^i = - \zeta^i(t,\vec{x})$.\footnote{The signs of the FDiff transformations are the opposite of the standard diffeomorphisms.} The FDiff transformations on the ADM variables has the following form:
\begin{eqnarray}
 &&
 \delta N = 
 \zeta^{k} \partial_{k} N + f \dot{N} + \dot{f}N \,, 
 \label{FdiffN}
 \\ &&
 \delta N^i = 
 \zeta^k \partial_k N^i - N^k \partial_k \zeta^i + \dot{\zeta}^i 
 + f \dot{N}^{i} + \dot{f} N^i \,,
 \label{FdiffNi}
 \\ &&
 \delta g_{ij} = 
 \zeta^{k} \partial_{k} g_{ij} + 2g_{k(i}\partial_{j)} \zeta^{k} 
 + f \dot{g}_{ij} \,.
 \label{Fdiffg}
\end{eqnarray}
These general FDiff transformations are a feature of the Ho\v{r}ava theory at the classical level. In the quantization of the theory we impose asymptotically flat boundary conditions, by fixing specific values of the fields at spatial infinity. The fixed values are equivalent to have set a Cartesian system at spatial infinity (see Ref.~\cite{Bellorin:2012di} for a discussion on asymptotically flat conditions in the context of Ho\v{r}ava theory, where a definition of the spacetime metric does not exist). These conditions on the spatial metric and the lapse function are
\begin{equation}
 g_{ij}  = \delta_{ij} + \mathcal{O}(r^{-1}) \,,
 \quad
 N = 1 + \mathcal{O}(r^{-1}) \,,
 \quad
 r \equiv \sqrt{x^i x^i} 
\end{equation} 
(The field $N^i$ remains as a Lagrange multiplier). Since $N|_{r=\infty} = 1$ and $f(t)$ is independent of the spatial point, the boundary condition requires the restriction $f(t) = 0$. For example, the terms $\dot{f}(t) N + f(t) \dot{N}$ in (\ref{FdiffN}) break this condition on $N$ if $f$ is left active. Notice that the way to impose the restriction $f=0$ is equivalent to a partial gauge-fixing condition since we are recurring to specific values of the fields associated with a coordinate system, at least at infinity. Therefore, the FDiff gauge symmetry of the quantum theory is characterized by the time-dependent spatial vector $\zeta^i(t,\vec{x})$. We denote these transformations by $\delta_\zeta$; they are given by
\begin{eqnarray}
&&
\delta_\zeta N = 
\zeta^{k} \partial_{k} N \,, 
\label{FdiffNrestricted}
\\ &&
\delta_\zeta N^i = 
\zeta^k \partial_k N^i - N^k \partial_k \zeta^i + \dot{\zeta}^i \,,
\label{FdiffNirestricted}
\\ &&
\delta_\zeta g_{ij} = 
\zeta^{k} \partial_{k} g_{ij} + 2g_{k(i}\partial_{j)} \zeta^{k} \,.
\label{Fdiffgrestricted}
\end{eqnarray}
It is important to compare with the spatial diffeomorphisms, since many variables behave as spatial tensors that evolve in time, as the case of $N$, $g_{ij}$, and the arbitrary parameter $\zeta^i$ itself. For the case of a time-dependent spatial tensor $T^{ij\cdots}$ and a tensor density $t^{ij\cdots}$, their FDiff transformations are functionally identical to spatial diffeomorphism:
\begin{equation}
 \begin{array}{l}
 \delta_\zeta T^{ij\cdots} =
 \zeta^k \partial_k T^{ij\cdots} - T^{kj\cdots} \partial_k \zeta^i 
 - T^{ik\cdots} \partial_k \zeta^j 
 - \cdots
 \,,
 \\
 \delta_\zeta t^{ij\cdots} =
   \zeta^k \partial_k t^{ij\cdots} + t^{ij\cdots} \partial_k \zeta^k
 - t^{kj\cdots} \partial_k \zeta^i 
 - t^{ik\cdots} \partial_k \zeta^j 
 - \cdots   
 \,.
 \end{array} 
 \label{tensors}
\end{equation}
Throughout this study, the term FDiff gauge symmetry of the Ho\v{r}ava theory refers to the transformations (\ref{FdiffNrestricted}) -- (\ref{Fdiffgrestricted}), and (\ref{tensors}) for the case of time-dependent spatial tensors/densities. Among the ADM variables, only the shift vector $N^i$ has a FDiff transformation that is functionally different from	 a spatial diffeomorphism.

The classical action of the nonprojectable theory is \cite{Horava:2009uw,Blas:2009qj}
\begin{equation}
S = \int dt d^3x \sqrt{g} N \left( K_{ij} K^{ij} - \lambda K^2 - \mathcal{V} \right) \,,
\label{classicalaction}
\end{equation}
where the kinetic terms are defined in terms of the extrinsic curvature tensor,
\begin{equation}
K_{ij} = \frac{1}{2N} \left( \dot{g}_{ij} - 2 \nabla_{(i} N_{j)} \right) \,,
\qquad
K = g^{ij} K_{ij} \,.
\end{equation}
$\mathcal{V} = \mathcal{V}[g_{ij},a_i]$ is called the potential, where 
\begin{equation}
a_i = \frac{\partial_i N}{N} \,.
\end{equation} 
$\mathcal{V}$ contains all the terms with spatial derivatives that are compatible with the FDiff gauge symmetry, including the higher-order ones characteristic of the Ho\v{r}ava theory. The order of each term is labeled by the parameter $z$, such that a given term is of $2z$ order in spatial derivatives. The power-counting criterion for renormalizability requires terms of order $z=3$ for the theory in 3 spatial dimensions. Although the complete potential is huge, for the study of renormalization we may concentrate on a certain class of terms explicitly. The reason is that the only ingredients we require to know explicitly are the propagators. Regarding the vertices, we are only required to know the possible orders that they contribute in spatial derivatives, which has been fixed at the beginning (besides some restrictions on the couplings of certain fields, which we have studied previously \cite{Bellorin:2022qeu,Bellorin:2023dwk}). Moreover, in the propagators we are interested in only the dominant contributions at the ultraviolet; that is, the $z=3$ terms. Therefore, in this study we consider explicitly only the $z=3$ terms of the potential that contribute to the propagators \cite{Colombo:2014lta}. In this way, the terms of the potential that we handle explicitly are\footnote{The coupling constants of the theory are $\lambda, \alpha_{3}, \alpha_{4}, \beta_{3}, \beta_{4}$. We use the shorthand $\bar{\sigma}=\lambda/(1-3\lambda)$.}
\begin{equation}
\mathcal{V}^{\text{($z=3$ propag)}} =
- \alpha_{3}\nabla^{2}R\nabla_{i}a^{i}
- \alpha_{4}\nabla^{2}a_{i}\nabla^{2}a^{i}
- \beta_{3}\nabla_{i}R_{jk}\nabla^{i}R^{jk}
- \beta_{4}\nabla_{i}R\nabla^{i}R \,.
\label{potencial} 
\end{equation}
$\nabla_i$ and $R_{ij}$ are the covariant derivative and the Ricci tensor of the spatial metric $g_{ij}$.

The primary classical Hamiltonian of the nonprojectable Ho\v{r}ava theory is given by \cite{Kluson:2010nf,Donnelly:2011df,Bellorin:2011ff}
\begin{equation}
H_0 =
\int d^3x \mathcal{H}_0 \,,
\quad
\mathcal{H}_0 \equiv 
\sqrt{g} N \left( \frac{\pi^{ij}\pi_{ij}}{g} 
+ \bar{\sigma} \frac{\pi^{2}}{g} + \mathcal{V} \right) \,.
\label{H0}
\end{equation}
The classical canonical conjugate pairs are $(g_{ij},\pi^{ij})$ and $N$ with its conjugate momentum. We denote the trace $\pi \equiv g^{ij} \pi_{ij}$. The canonical momentum of $N$ is zero due to the constraints of the theory; we discard it from the phase space. 

The BFV quantization requires one to identify the constraints that are involutive under Dirac brackets \cite{Fradkin:1975cq,Batalin:1977pb,Fradkin:1977xi}. In the case of the Ho\v{r}ava theory this is the momentum constraint 
\begin{equation}
\mathcal{H}_i =
- 2 g_{ij} \nabla_k \pi^{kj} = 0 \,,
\label{momentumconts}
\end{equation}
which is the generator of the spatial diffeomorphisms on the canonical pair $( g_{ij} , \pi^{ij} )$. The Dirac bracket between two $\mathcal{H}_i$ coincides with the Poisson bracket, yielding the algebra of spatial diffeomorphisms. By denoting two spatial points by $x^i$ and $y^i$ and leaving the time dependence implicit, this algebra is
\begin{equation}
\{\mathcal{H}_i(x),\mathcal{H}_j(y)\}_{\text{D}} =
\frac{\partial \delta(x-y) }{\partial x^i}\mathcal{H}_j(x)
- \frac{\partial \delta(x-y) }{\partial y^j}\mathcal{H}_i(y)\,,
\end{equation}
where $\{\,,\}_{\text{D}}$ indicates Dirac brackets. The coefficients of this algebra are used in the definition of the BRST charge and the gauge-fixed Hamiltonian.

The second-class constraints are given by the vanishing of the momentum conjugate to $N$, which we have already considered as solved, and the constraint
\begin{equation}
\theta_1 \equiv 
N \frac{ \delta H_0 }{ \delta N } 
=
\frac{N}{\sqrt{g}}\left(\pi^{ij}\pi_{ij}
+ \bar{\sigma}\pi^{2}\right)
+ \sqrt{g} N \mathcal{V}
+ \mathcal{B}
= 0 \,,
\label{theta2}
\end{equation}
where $\mathcal{B}$ stands for total derivatives.\footnote{Variations of $a_i$ with respect to $N$ produce a total derivative: $\delta a_i = \partial_i ( \delta N / N )$.} In the quantum theory, the terms in $\mathcal{B}$ that contribute to the propagators come from the terms (\ref{potencial}), such that 
\begin{equation}
 \mathcal{B}^{\text{($z=3$ propag})} = 
 - \alpha_{3}\sqrt{g} \nabla^{2}(N\nabla^{2}R)
 + 2\alpha_{4} \sqrt{g} \nabla^{i}\nabla^{2}(N\nabla^{2}a_{i}) \,.
\end{equation}
Notice that the difference between $\theta_1$ and the primary Hamiltonian density $\mathcal{H}_0$ (\ref{H0}) is the set of total derivatives $\mathcal{B}$. Therefore, the primary Hamiltonian (\ref{H0}) is equivalent to the integral of this second-class constraint, 
\begin{equation}
H_{0} = 
\int d^3x \, \theta_{1}
\,.
\label{primaryhamiltonian}
\end{equation}

\subsection{BFV quantization with second-class constraints and a gauge-fixing condition}
The BFV quantization adds the canonical pair $(N^{i}, \pi_{i})$ and the BFV ghost pairs $(C^{i},\bar{\mathcal{P}}_{i})$, $(\bar{C}_{i},\mathcal{P}^{i})$. The definition of the BFV path
integral, under a given gauge-fixing condition, is
\begin{eqnarray}
&&
Z =
\int \mathcal{D}V e^{iS} \,,
\label{Z}
\\
&&
\mathcal{D}V = 
\mathcal{D} g_{ij} \mathcal{D}\pi^{ij} \mathcal{D}N \mathcal{D} N^{k}\mathcal{D}\pi_{k}
\mathcal{D} C^i \mathcal{D} \bar{\mathcal{P}}_i
\mathcal{D} \bar{C}_i \mathcal{D} \mathcal{P}^i
\times \delta(\theta_{1}) \det\frac{ \delta \theta_1 }{ \delta N } \,, \;\;\;
\label{measure}
\\ &&
S=
\int dt d^3x \left( 
\pi^{ij} \dot{g_{ij}} 
+ \pi_i \dot{N}^i 
+ \bar{\mathcal{P}}_{i} \dot{C}^{i} 
+ \mathcal{P}^{i} \dot{\bar{C}}_{i} 
- \mathcal{H}_{\Psi}
\right) \,.
\end{eqnarray}
The factor $\delta(\theta_1) \det \delta\theta_1/\delta N$ in (\ref{measure}) is the measure associated with the second-class constraints, see Ref.~\cite{Bellorin:2022efu}. It can be incorporated in the quantum action with the help of new auxiliary variables,
\begin{equation}
\delta(\theta_{1}) \det \frac{ \delta \theta_1 }{ \delta N } =	
\int\mathcal{D}\mathcal{A} 
\mathcal{D}\bar{\eta} \mathcal{D}\eta
\exp\left[
i \int dt d^3x\left(
\mathcal{A}\theta_1
- \bar{\eta}  \frac{ \delta \theta_1 }{ \delta N } \eta
\right)
\right] \,,
\label{measureA}
\end{equation}
where $\mathcal{A}$ is a bosonic scalar field and $\eta,\bar{\eta}$ is a pair of scalar ghosts. The gauge-fixed Hamiltonian density is defined by
\begin{equation}
\mathcal{H}_{\Psi} =
\mathcal{H}_0 
+ \{\Psi,\Omega\}_{\text{D}} \,,
\label{bfvhamiltonian}
\end{equation}
where $\Omega$ is the generator of the BRST symmetry, given by
\begin{equation}
\Omega =
\int\,d^3x
\left(
\mathcal{H}_kC^k
+\pi_k\mathcal{P}^k
-C^k\partial_kC^l\bar{\mathcal{P}}_l
\right)\,,
\end{equation}
and $\Psi$ is a gauge fermion.

For the gauge fixing we may use the original BFV structure of the gauge fermion \cite{Bellorin:2021udn,Bellorin:2021tkk}. The gauge-fixing condition and its associated gauge fermion are, respectively, of the form
\begin{eqnarray}
&&
\dot{N}^i - \chi^i = 0\,,
\label{relativisticgaugephi}
\\ &&
\Psi =
\bar{\mathcal{P}}_{i} N^{i} + \bar{C}_{i} \chi^{i} \,,
\label{relativisticgaugepsi}
\end{eqnarray} 
where $\chi^i$ is a part that must be chosen. With these settings, the BFV path integral of the Ho\v{r}ava theory takes the form
\begin{equation}
\begin{split}
Z =& 
\int \mathcal{D} g_{ij} \mathcal{D}\pi^{ij} \mathcal{D}N \mathcal{D} N^{k}\mathcal{D}\pi_{k}
\mathcal{D} C^i \mathcal{D} \bar{\mathcal{P}}_i
\mathcal{D} \bar{C}_i \mathcal{D} \mathcal{P}^i
\mathcal{D}\mathcal{A} 
\mathcal{D}\bar{\eta} \mathcal{D}\eta
\\ &
\exp\left[ i \int dt d^3x 
\Big(
\pi^{ij} \dot{g_{ij}} 
+ \pi_i \dot{N}^i 
+ \bar{\mathcal{P}}_{i} \dot{C}^{i} 
+ \mathcal{P}^{i} \dot{\bar{C}}_{i}
- \mathcal{H}_{0} 
- \mathcal{H}_i N^i
- \bar{\mathcal{P}}_i \mathcal{P}^i 
\right. 
\\ &  \left.
+ \bar{\mathcal{P}}_i \left( C^j \partial_j N^i - N^j \partial_j C^i  \right)
- \pi_k\chi^{k}
- \bar{C}_i \{\chi^{i} \,, \mathcal{H}_j\} C^j
+ \mathcal{A}\theta_1
- \bar{\eta}  \frac{ \delta \theta_1 }{ \delta N } \eta
\Big)\right]
\,.
\end{split}
\label{quantumaction} 
\end{equation}
To arrive at this form of the quantum action, we have used the fact that the factor $\chi^i$ we choose does not depend on $N^i$ nor any of the ghost fields.

To write $\chi^i$ explicitly, we introduce perturbative variables around a flat background: $g_{ij} = \delta_{ij}$, $N =1$, and the rest of variables take zero value. The perturbations are denoted by
\begin{equation}
g_{ij} = \delta_{ij} + h_{ij} \,,
\quad
\pi^{ij} = p^{ij} \,,
\quad
N = 1 + n \,,
\quad
N^i = n^i \,,
\end{equation}
and for the rest of the perturbative field variables we keep the original notation. We choose $\chi^{i}$ to be the local expression\footnote{The coefficients in (\ref{gaugefixing}) are chosen to simplify the resulting propagators (see Refs.~\cite{Barvinsky:2015kil,Bellorin:2021udn}).}
\begin{equation}
\chi^{i} = 
\rho\mathfrak{D}^{ij}\pi_{j}
- 2 \rho \Delta^{2}\partial_{j}h_{ij}
+ 2 \rho \lambda \bar{\kappa} \Delta^{2}\partial_{i}h
- 2 \kappa \rho \Delta\partial_{i}\partial_{j}\partial_{k}h_{jk}  \,, 
\label{gaugefixing}
\end{equation}
where 
\begin{equation}
\mathfrak{D}^{ij} = \delta_{ij} \Delta^2 
+ \kappa \Delta \partial_i \partial_j \,, 
\end{equation}
$\Delta = \partial_k \partial_k$, $\rho,\kappa$ are independent constants, and $\bar{\kappa} = \kappa + 1$. The explicit form of $\chi^i$ completes the procedure of the gauge fixing.

\section{The BRST-symmetry structure}
In the BFV formalism, the BRST symmetry transformations on the canonical fields are generated by $\Omega$, according to the rule
\begin{equation}
\delta_\Omega \Phi = 
\{ \Phi \,, \Omega \}_{\text{D}}\, \epsilon \,,
\label{brstgeneral}
\end{equation}
where $\epsilon$ is the fermionic parameter of the transformation. The BRST transformations of the quantum canonical fields, when the theory is restricted to the constrained phase space (the constrained surface), result in
\begin{equation}
\begin{array}{ll}
\delta_\Omega g_{ij} =
\delta_{C\epsilon} g_{ij} \,,
&
\delta_\Omega \pi^{ij} =
\delta_{C\epsilon} \pi^{ij} \,,
\\[1ex]
\delta_\Omega N =
\delta_{C\epsilon} N \,, 
\qquad
& \\[1ex]
\delta_\Omega N^i = 
\mathcal{P}^i \epsilon \,,
&
\delta_\Omega \pi_i = 0 \,,
\\[1ex]
\delta_\Omega C^{i} = 
\partial_j C^i C^j \epsilon \,,
& 
\delta_\Omega \bar{\mathcal{P}}_{i} =
\delta_{C\epsilon} \bar{\mathcal{P}}_i  \,, 
\\[1ex]
\delta_\Omega \mathcal{P}^i = 0 \,,
&
\delta_\Omega \bar{C}_i = 
\pi_i \epsilon \,,
\end{array}
\label{brst}
\end{equation}
where $\delta_{C\epsilon}$ is the FDiff with the time-dependent vector parameter $C^i \epsilon$. $\pi^{ij}$ and $\bar{\mathcal{P}}_i$ are time-dependent spatial tensor densities. 

The auxiliary fields $\mathcal{A}, \eta,\bar{\eta}$ are not canonical. We define their BRST transformation in such a way that the measure is left invariant. The transformations are FDiff along $C^i\epsilon$:
\begin{equation}
\begin{split}
&
\delta_\Omega \mathcal{A} = 
\delta_{C\epsilon} \mathcal{A} \,,
\\ &
\delta_\Omega \eta = 
\delta_{C\epsilon} \eta  \,,
\\ &
\delta_\Omega \bar{\eta} = 
\delta_{C\epsilon} \bar{\eta}  \,,
\end{split}
\label{brstsecondclass}
\end{equation}
where the three fields $\mathcal{A},\eta,\bar{\eta}$ transform as time-dependent spatial scalar fields. We remark that $\mathcal{A}$ multiplies the constraint $\theta_1$, which is invariant under the BRST transformation (\ref{brstgeneral}) since it is a second-class constraint. As a consequence the combination $\mathcal{A} \theta_1$ is invariant over the constrained surface.\footnote{Alternatively, one may declare $\mathcal{A}$ to be BRST invariant. In this case the combination $\mathcal{A}\theta_1$ is left invariant in the whole phase space, not only on the constrained surface.} The factor $\delta \theta_1 / \delta N$ multiplying $\eta,\bar{\eta}$ is a time-dependent scalar density.

On the path integral (\ref{quantumaction}), we perform the integration on the ghost fields $\mathcal{P}^i,\bar{\mathcal{P}}_i$. The resulting action can be grouped into two sectors,
\begin{equation}
  S = S_0[\varphi^a] + S_\Omega \,,
  \label{quantumactionfinal}
\end{equation}
where
\begin{eqnarray}
  &&
  S_0[\varphi^a] =
  \int dt d^3x \left( 
  \pi^{ij} \dot{g_{ij}} 
  - \mathcal{H}_0 
  - N^i \mathcal{H}_i
  + \mathcal{A}\theta_1
  - \bar{\eta}  \frac{ \delta \theta_1 }{ \delta N } \eta
  \right) \,,
  \label{Sgauge}
  \\ &&
  S_\Omega  =
  \int dt d^3x 
    \left[ \pi_i \left( \dot{N}^i - \chi^i \right) 
  - \dot{\bar{C}}_i \left( 
       \dot{C}^i + C^j \partial_j N^i - N^j \partial_j C^i  \right)
  - \bar{C}_i \{ \chi^i , \mathcal{H}_j \} C^j 
  \right] \,.
  \nonumber \\
  \label{SOmega}
\end{eqnarray}
$S_0[\varphi^a]$ depends exclusively on the set of fields $\varphi^a$, where 
\begin{equation}
\varphi^a = \{ g_{ij},\pi^{ij},N,N^i,\mathcal{A},\eta,\bar{\eta} \} \,.
\label{representationR}
\end{equation}
At this point it is useful to clarify that all the fields of the quantum theory transform as time-dependent spatial tensors/densities under FDiff transformations, except for $N^i$. Moreover, $S_0$ is invariant under arbitrary FDiff gauge transformations. We remark that, for a FDiff transformation with a time-dependent vector parameter $\zeta^i$, the first and third terms of (\ref{Sgauge}) combine themselves to cancel time derivatives of $\zeta^i$,
\begin{equation}
 \delta_\zeta 
  \int dt d^3x \left( 
    \pi^{ij} \dot{g_{ij}} - N^i \mathcal{H}_i \right) 
 = 0 \,,
 \label{invarianceNi}
\end{equation}
as it is well known from the ADM formulation of general relativity. The rest of the terms in (\ref{Sgauge}) contain no time derivatives and are independent of $N^i$. Their invariance under FDiff is automatic since they are written totally in terms of spatial tensors/densities. In contrast, $S_\Omega$ is the gauge-fixing sector of this symmetry.

As a consequence of the previous integration, the BRST symmetry (\ref{brst}) must be revised. Specifically, the transformation of $N^i$ is affected. The second term in (\ref{SOmega}) is key to unfold the new transformation, since it has the form of a FDiff (\ref{FdiffNirestricted}) along $C^i$. Therefore, we define the new BRST transformation of $N^i$ to be the FDiff
\begin{equation}
 \delta_\Omega N^i =
 \left( C^j \partial_j N^i - N^j \partial_j C^i + \dot{C}^i \right) \epsilon \,.
\end{equation}
This transformation is nilpotent. The new BRST symmetry transformations are
\begin{equation}
\begin{array}{llll}
  \delta_\Omega g_{ij} =
  \delta_{C\epsilon} g_{ij} \,,
  &
  \delta_\Omega \pi^{ij} =
  \delta_{C\epsilon} \pi^{ij} \,,
  &
  \delta_\Omega N =
  \delta_{C\epsilon} N \,, 
  &
  \delta_\Omega N^i = 
  \delta_{C\epsilon} N^i \,,
  \\[1ex]
  \delta_\Omega \mathcal{A} = 
  \delta_{C\epsilon} \mathcal{A} \,,
  &
  \delta_\Omega \eta = 
  \delta_{C\epsilon} \eta  \,,
  &
  \delta_\Omega \bar{\eta} = 
  \delta_{C\epsilon} \bar{\eta}  \,,
  \\[1ex]
  \delta_\Omega C^{i} = 
  \partial_j C^i C^j \epsilon \,,
  & 
  \delta_\Omega \bar{C}_i = 
  \pi_i \epsilon \,,
  &
  \delta_\Omega \pi_i = 0 \,.
\end{array}
\label{brstnew}
\end{equation}
After the (re)definitions we have done, it turns out that the BRST transformation of all the $\varphi^a$ fields corresponds to a FDiff along $C^i\epsilon$. Therefore, the BRST invariance of $S_0[\varphi^a]$ is automatic.

The quantum action (\ref{quantumactionfinal}) can be written in the standard notation of the BRST symmetry. We denote by $\boldsymbol{s}$ the BRST operator. The action of $\boldsymbol{s}$ on the $\varphi^a$ fields is a FDiff transformation with a vector parameter equal to $C^i$. $\bar{C}_i$ and $\pi_i$ are the usual auxiliary fields of the BRST symmetry. The action of the BRST operator is
\begin{equation}
\begin{split}
&
\boldsymbol{s} \varphi^a = \delta_C \varphi^a \,,
\\ &
\boldsymbol{s} C^i = 
- C^j \partial_j C^i \,,
\\ &
\boldsymbol{s} \bar{C}_i = 
\pi_i \,,
\\ &
\boldsymbol{s} \pi_i = 0 \,.
\end{split}
\label{brsts}
\end{equation}
The sector $S_\Omega$ (\ref{SOmega}) is equal to the action of the BRST operator on a gauge fermion, $S_\Omega = \int \boldsymbol{s} \tilde{\Psi}$, where
\begin{equation}
\tilde{\Psi} = 
\bar{C}_i \left( \dot{N}^i - \chi^i \right) \,,
\label{gaugefermionpre}
\end{equation}
and $\chi^i$ is given in (\ref{gaugefixing}). Therefore, the quantum action (\ref{quantumactionfinal}) has the BRST-invariant form
\begin{equation}
S = 
S_0[\varphi^a] + \int dt d^3x \boldsymbol{s} \tilde{\Psi} \,. 
\label{quantumactionbrst}
\end{equation}
We highlight that the whole Lagrangian is completely local. 

\section{Propagators and locality of divergences}
The propagators can be calculated from the action (\ref{quantumactionfinal}), expanded at second order in perturbations. In Fourier space $(\omega,\vec{k})$, after a Wick rotation, the nonzero propagators are
\begin{align}
& 
\langle \bar{C}_{i}C^j\rangle = 
- 2P_{ij}\mathcal{T}_{3} 
- 2\hat{k}_{i}\hat{k}_{j}\mathcal{T}_{4}\,,
\nonumber 
\\ &
\langle p^{ij}p^{kl}\rangle =
- \beta_{3}M_{ijkl}k^{6}\mathcal{T}_{1}
- 2\nu P_{ij}P_{kl} k^{6} \mathcal{T}_{2}\,,
\nonumber 
\\ &
\langle h_{ij}h_{kl}\rangle =
4M_{ijkl}\mathcal{T}_{1} 
+ 8\left(\omega^2+2\rho k^6\right)Q_{ijkl}\mathcal{T}_{3}^{2}
+ 8\left(\sigma P_{ij}P_{kl}
+ \bar{\sigma}P_{ij}\hat{k}_{k}\hat{k}_{l} 
+ \bar{\sigma}\hat{k}_{i}\hat{k}_{j}P_{kl}\right) \mathcal{T}_2
\nonumber
\\&
+ \frac{4}{1-\lambda}\hat{k}_{i}\hat{k}_{j}\hat{k}_{k}\hat{k}_{l}\left[\left(2\omega^2\mathcal{T}_4 - 1\right)\mathcal{T}_4
+ 2 \lambda\bar{\sigma} \mathcal{T}_2 \right]\,,
\nonumber 
\\ &
\langle h_{ij}p^{kl}\rangle = 
2\omega \left[M_{ijkl}\mathcal{T}_{1}
+ 2Q_{ijkl}\mathcal{T}_{3}
+ 2 P_{ij}P_{kl}\mathcal{T}_{2}
+ 2\bar{\sigma}\hat{k}_{i}\hat{k}_{j}P_{kl}\left(\mathcal{T}_2-\mathcal{T}_4\right)
+ \hat{k}_{i}\hat{k}_j\hat{k}_{k}\hat{k}_{l}\mathcal{T}_{4}\right]\,,
\nonumber 
\\ &
\langle n^kh_{ij}\rangle =
- 16i\rho \omega k^{4}\left(P_{k(i}k_{j)}\mathcal{T}_{3}^2
+ \bar{\kappa} k_k\hat{k}_i\hat{k}_{j} \mathcal{T}_4^2\right)\,,
\nonumber
\\ &
\langle n^ip^{kl}\rangle =
- 4i\rho k^{4} P_{i(k}k_{l)}\mathcal{T}_3
+ 4i\rho\bar{\kappa}\left(2\lambda P_{kl}
- (1-\lambda)\hat{k}_{k}\hat{k}_{l}\right)k_{i}k^{4}\mathcal{T}_{4}\,,
\nonumber
\\ &
\langle n^in^j\rangle = 
4\rho\left(\omega^{2} + 2\rho k^{6}\right) k^{4}P_{ij} \mathcal{T}^{2}_3  
+ 4\rho\bar{\kappa}\left(\omega^2+4\rho\bar{\kappa}(1-\lambda)k^6\right) k^{4} \hat{k}_i\hat{k}_{j}\mathcal{T}^{2}_4 \,,
\nonumber 
\\ &
\langle nh_{ij}\rangle = 
- \frac{4\alpha_{3}}{\alpha_{4}}\left(
\sigma P_{ij}
+ \bar{\sigma} \hat{k}_i\hat{k}_j\right)\mathcal{T}_{2}\,,
\nonumber 
\\ &
\langle np_{ij}\rangle = 
- \frac{2\alpha_3}{\alpha_4}\omega P_{ij}\mathcal{T}_2\,,
\nonumber
\\ &
\langle \pi_kn^j\rangle =
2\omega\left(\mathcal{T}_3
P_{kj} 
+ \hat{k}_k\hat{k}_j\mathcal{T}_4\right)\,,
\nonumber
\\&
\langle \pi_kh_{ij}\rangle = 
- 4i\left(P_{i(j}k_{k)}\mathcal{T}_3 
+ k_k\hat{k}_i\hat{k}_j\mathcal{T}_4\right)\,,
\nonumber
\\&
\langle nn\rangle = 
\frac{2\alpha_{3}^{2}}{\alpha_{4}^{2}} \sigma\mathcal{T}_2 \,,
\nonumber \\
\label{propagators}
\end{align}
and
\begin{equation}
\langle\mathcal{A}\mathcal{A}\rangle = 
\langle \mathcal{A}n\rangle = 
\langle \bar{\eta}\eta\rangle =
- \frac{1}{\alpha_{4}k^{6}}\,,
\label{irregularprop}
\end{equation}
where the projectors are defined by
\begin{equation}
\begin{split}
&
P_{ij}  = 
\frac{1}{2}(\delta_{ij}-\hat{k}_i\hat{k}_j)\,,
\\ &
M_{ijkl} = P_{ik}P_{jl} + P_{il}P_{jk} - P_{ij}P_{kl}\,,
\\ &
Q_{ijkl} = \hat{k}_{i}\hat{k}_{k}P_{jl}
+ \hat{k}_{j}\hat{k}_{k}P_{il}
+ \hat{k}_{i}\hat{k}_{l}P_{jk}
+ \hat{k}_{j}\hat{k}_{l}P_{ik}\,,
\end{split}
\end{equation}
and
\begin{align}
&
\mathcal{T}_1 = \frac{1}{\omega^{2}+\beta_{3}k^{6}} \,, 
\qquad
\mathcal{T}_2 =  \frac{1}{\omega^{2}
	+ \sigma\nu k^{6}} \,,
\nonumber \\ &
\mathcal{T}_3 = \frac{1}{\omega^{2}-2\rho k^{6}} \,,
\qquad
\mathcal{T}_4 = \frac{1}{\omega^{2} - 4\rho\bar{\kappa}(1-\lambda)k^{6}} \,,
\\ &
\nu = 3\beta_{3} + 8 \beta_{4} - 2 \frac{\alpha_{3}^{2}}{\alpha_{4}} \,,
\qquad 
\sigma=\frac{1-\lambda}{1-3\lambda} \,.
\end{align}

The condition of regularity on propagators is appropriate for the study of ultraviolet divergences in Lorentz-violating theories \cite{Anselmi:2008bq}.\footnote{Throughout this study we assume that infrared divergences have been regularized.} Indeed, the gauge condition (\ref{gaugefixing}) is intended to get regular propagators for the quantum fields \cite{Barvinsky:2015kil,Bellorin:2021udn}. Consider a propagator between two fields that have scaling dimensions $r_1$ and $r_2$. It is regular if it is given by the sum of terms of the form
\begin{equation}
\frac{ P(\omega,k^i) }{ D(\omega,k^i) } \,,
\label{defregular}
\end{equation}
where $D$ is the product
\begin{equation}
D = 
\prod_{m=1}^{M} ( A_m \omega^2 + B_m k^{2d} + \cdots ) \,, 
\quad
A_m \,, B_m > 0 \,,
\label{regular}
\end{equation}
and $P(\omega,k^i)$ is a polynomial of maximal scaling degree less than or equal to $r_1 + r_2 + 2(M-1)d$, with $d=3$ in our case. The ellipsis stands for terms with lower scaling. All the propagators in the list (\ref{propagators}) are given in terms of sums of products of the four factors $\mathcal{T}_1, \mathcal{T}_2, \mathcal{T}_3, \mathcal{T}_4$, and these propagators satisfy the condition of regularity, whenever the constants satisfy the following bounds:
\begin{equation}
\beta_3 > 0 \,, 
\quad
\sigma\nu > 0 \,, 
\quad 
\rho < 0\,, 
\quad 
( 1 - \lambda ) \bar{\kappa}  > 0  \,.
\end{equation}
On the contrary, the three propagators $\langle \mathcal{A} \mathcal{A} \rangle$,  $\langle \mathcal{A} n \rangle$, and $\langle \bar{\eta}\eta \rangle$ in (\ref{irregularprop}), which are the only ones involving the fields  $\mathcal{A},\eta,\bar{\eta}$, are independent of $\omega$; hence these propagators are irregular. The persistence of irregular propagators in the nonprojectable theory demands a careful study of the divergences. We remark that this effect is a consequence of the measure of the second-class constraints.

In the action (\ref{quantumactionfinal}), time derivatives arise uniquely in terms that are of second order in perturbations. As a consequence, vertices do not depend on the frequency $\omega$. Hence, for the integration on $\omega$ we only need to consider propagators. We call an irregular loop to a loop formed completely with the irregular propagators (\ref{irregularprop}). Since these propagators do not depend on $\omega$, an irregular loop produces a divergence of the kind $\sim \int d\omega$. Such a divergence multiplies any diagram containing (at least) one irregular loop. In previous analysis \cite{Bellorin:2022qeu,Bellorin:2023dwk}, we have shown that all the diagrams with irregular loops cancel completely between them. We may give a clue on how this happens in the effective action. The integrated form of the one-loop quantum corrections to the effective action has the form of a Berezinian,
\begin{equation}
	i \Gamma^{(1)} =
	- \frac{1}{2} \ln \left( 
	\frac{ \det ( \mathbb{A} - 2 \mathbb{B} \mathbb{D}^{-1} \mathbb{B}^T ) }
	{ \det \mathbb{D} } \right)  \,,
	\label{Berfull}
\end{equation}
where $\mathbb{A},\mathbb{B},\mathbb{D}$ are matrices of derivatives of the action. We remark that all the irregular propagators in (\ref{irregularprop}) are exactly equal among them. Some restrictions on the possible couplings of the $\mathcal{A},\eta,\bar{\eta}$ fields are required in this analysis; they are summarized in Ref.~\cite{Bellorin:2023dwk}. For example, an important fact is that there is no way to form an irregular loop mixing propagators of $\mathcal{A}$ with the propagator of $\eta,\bar{\eta}$. Irregular loops containing the $\mathcal{A}$ field are produced exclusively by the numerator of (\ref{Berfull}), whereas irregular loops containing the $\eta,\bar{\eta}$ ghosts are produced exclusively by the factor $\det \mathbb{D}$ in the denominator. They cancel themselves exactly, and there are no more irregular loops in the effective action (\ref{Berfull}). 

Moreover, only the irregular loops produce divergences on the integration on $\omega$, due to the fact that the regular propagators automatically render the integration on $\omega$ finite. Let us suppose first a loop composed completely of regular propagators. The regular propagators with the lowest scaling in $\omega^{-1}$ are of order $\sim \omega^{-1}$. These are the propagators $\langle h_{ij} p^{kl} \rangle$, $\langle n p_{ij} \rangle$, and $\langle \pi_k n^j \rangle$. If the loop consists only of one propagator of this kind, then the integral is zero since these propagators are odd in $\omega$. The next order is a product of two propagators of this kind, or a single propagator with scaling $\sim \omega^{-2}$. In both cases the integral in $\omega$ is finite.\footnote{This reasoning is not altered by considering external frequency circulating in some step of the loop.} By increasing the number of regular propagators, the convergence in the integration on $\omega$ becomes faster. Now consider the presence of one or more irregular propagators (\ref{irregularprop}) in the loop, but not all since we know that irregular loops cancel completely. Since the irregular propagators are independent of $\omega$, the analysis of the integration on $\omega$ is identical to the previous case of a loop made exclusively of regular propagators.

In the integration on the spatial momentum $k^i$, all the propagators have a regular structure on this variable, including the ones of the fields $\mathcal{A},\eta,\bar{\eta}$. According to the analysis of Lorentz-violating theories \cite{Anselmi:2007ri,Anselmi:2008bq}, the locality of divergences produced by the integration on $k^i$ is ensured. 

On the basis of the scaling of propagators and the maximal number of spatial derivatives in the vertices, we may compute the superficial degree of divergence $D_{\text{div}}$. For an arbitrary diagram (that may be a subdiagram), the result is
\begin{equation}
  D_{\text{div}} = 6- 3E_p - 2E_N - E_\pi - X\,,
\end{equation}
where $E_p$ is the number of external $p^{ij}$-legs, $E_N$ the number of external $n^i$-legs, $E_\pi$ the number of external $\pi_i$-legs, and $X$ the total number of spatial derivatives on external legs. The diagrams with the highest divergence have $D_{\text{div}} = 6$. This order is equal to the order of the bare Lagrangian, in agreement with the power-counting criterion used in the formulation of the classical theory.

\section{The background-field approach}
The aim of introducing background fields is to get a background-gauge symmetry in the gauge-fixed quantum action. This symmetry transforms simultaneously the quantum fields $\varphi^a$ and the background fields $\phi^a$ in the form of the original FDiff gauge transformations (\ref{FdiffNrestricted}) -- (\ref{tensors}), with the same parameter for both classes of fields. Specifically, one is required to handle the subset of fields $\varphi^a$ involved in the gauge-fixing condition, in terms of the linear combination $\varphi^a - \phi^a$. In our case, we require the introduction of background fields only for $g_{ij}$ and $N^i$, which we denote by $\bar{g}_{ij}$ and $\bar{N}^i$, respectively (hence, $\phi^a = \{ \bar{g}_{ij}, \bar{N}^i \}$). We use a notation for the difference of fields:\footnote{Not to confuse $h_{ij},n^i$ with the variables of section 2.} 
\begin{equation}
 h_{ij} =  g_{ij} - \bar{g}_{ij}  \,,
 \quad 
 n^i = N^i - \bar{N}^i  \,.
 \label{differences}
\end{equation}  
Because of the linearity of the gauge transformations on the parameter and the fields, $h_{ij}$ and $n^i$ transform exactly as time-dependent spatial tensors under background-gauge transformations,
\begin{eqnarray}
&&
\delta_\zeta h_{ij} = 
\zeta^k \partial_k h_{ij} + 2 h_{k(i} \partial_{j)} \zeta^k \,,
\\ &&
\delta_\zeta n^i = 
\zeta^k \partial_k n^i - n^k \partial_k \zeta^i \,.
\end{eqnarray}

The sector $S_0[\varphi^a]$ is unaltered in this procedure; hence, it is automatically invariant under background-gauge transformations. The gauge fermion (\ref{gaugefermionpre}) is replaced by a background-dependent one,
\begin{equation}
\begin{split}
\Psi_{\text{bg}} = \,&
\bar{C}_i \left( 
D_t n^i 
- \rho \Theta^{ijk} h_{jk}
- \rho \mathcal{D}^{ij} (\pi_j / \sqrt{ \bar{g} }) \right) 
- \mathbb{T}^{ij} h_{ij}
- \mathbb{K}_{ij} \pi^{ij}
- \mathbb{T}_i n^i
\\[1ex] &
- \mathbb{T} N
- \mathbb{S} \mathcal{A}
- \bar{\mathbb{N}} \eta
- \bar{\eta} \mathbb{N}
+ \bar{J}_i C^i \,,
\end{split}
\label{psibackground}
\end{equation}
where\footnote{In the definition of the operator $\Theta^{ijk}$ we have chosen the simplest combination of fifth-order covariant derivatives that reproduces the flat case (\ref{gaugefixing}). Unitarity requires the operator $\mathcal{D}^{ij}$ to be invertible.}
\begin{eqnarray}
&&
D_t n^i =
\dot{n}^i 
- \bar{N}^k \bar{\nabla}_k n^i
+ n^k\bar{\nabla}_k\bar{N}^i \,,
\\ &&
\Theta^{ijk} = 
- 2 \bar{g}^{ij} \bar{\nabla}^{4} \bar{\nabla}^k  
+ 2 \lambda \bar{\kappa} \bar{g}^{jk} \bar{\nabla}^{4}\bar{\nabla}^i
- 2\kappa \bar{\nabla}^2 \bar{\nabla}^i \bar{\nabla}^j \bar{\nabla}^k 
\,,
\\ &&
\mathcal{D}^{ij} = 
\bar{g}^{ij} \bar{\nabla}^4 
+ \kappa \bar{\nabla}^2 \bar{\nabla}^i \bar{\nabla}^j  
\,.
\label{Dbackground}
\end{eqnarray}
All indices are raised and lowered with the background metric $\bar{g}_{ij}$, and $\bar{\nabla}$ is its covariant derivative. In Eq.~(\ref{psibackground}) we have inserted external sources for the BRST transformations of the $(\varphi^a - \phi^a)$ fields. We denote these sources collectively by 
\begin{equation}
\gamma_a = 
\{ \mathbb{T}^{ij} , \mathbb{K}_{ij} , \mathbb{T}_i , \mathbb{T} , \mathbb{S} , \bar{\mathbb{N}} , \mathbb{N} \} \,, 
\end{equation}
whereas $\bar{J}_i$ is the source for $\boldsymbol{s} C^i$. Sources for the auxiliary fields $\bar{C}_i$ and $\pi_i$ are not included in $\Psi_{\text{bg}}$. All these sources transform as time-dependent spatial tensors/densities under FDiff. $D_t n^i$ transforms as a spatial vector under background-gauge transformations. The operators $\Theta^{ijk}$ and $\mathcal{D}^{ij}$ are made completely of spatial covariant derivatives; hence, $\Psi_{\text{bg}}$ is invariant under background-gauge transformations.

To write the action in the background-field approach, one introduces the 
operator 
\begin{equation}
\boldsymbol{Q} = 
\boldsymbol{s} + \Omega^a \frac{\delta}{\delta \phi^a} \,,
\end{equation}
where $\Omega^a = \{ \Omega_{ij} , \Omega^i \}$ are external Grassmann fields. $\boldsymbol{Q}$ is a nilpotent operator. The quantum gauge-fixed action in the presence of background fields takes the form
\begin{equation}
\Sigma_0 = 
S_0[\varphi^a] + \int dt d^3x\, \boldsymbol{Q} \Psi_{\text{bg}} \,.
\label{brststructure}
\end{equation}
More explicitly:
\begin{equation}
\begin{split}
 \Sigma_0 =&
   S_0
 + \int dt d^3x  \Big[
   \pi_i \left( 
   D_t n^i - \rho \Theta^{ijk} h_{jk}
   - \rho \mathcal{D}^{ij} ( \pi_j / \sqrt{\bar{g}} ) \right) 
 + \bar{C}_i \boldsymbol{s} \left( 
 D_t n^i - \rho \Theta^{ijk} h_{jk} \right) 
 \\ & 
 - \gamma_a \boldsymbol{s} \varphi^a
 + \bar{J}_i \boldsymbol{s} C^i 
 + \Omega^a \bar{C}_i 
    \frac{\delta}{\delta \phi^a} 
     \left( 
     D_t n^i - \rho \Theta^{ijk} h_{jk}
     - \rho \mathcal{D}^{ij} ( \pi_j / \sqrt{\bar{g}} ) \right) 
 + \Omega^a \gamma_a
 \Big] \,.
\end{split}
\label{explicitquantumactionbackground} 
\end{equation}

\section{Renormalization}
We collect the several results we have found here and in previous analysis: the BRST-invariant form (\ref{quantumactionbrst}) of the quantum action, together with its background-field extension (\ref{brststructure}); the completely local form of the gauge-fixed Lagrangian; the regularity of all the propagators that do not involve the fields $\mathcal{A},\eta,\bar{\eta}$; the cancellation of the irregular loops formed by the propagators of these fields; the absence of divergences along the $\omega$ direction, regardless of the presence of irregular propagators in diagrams; the regular structure of all the propagators with respect to the dependence on $k^i$, leading to the locality of the divergences; and the superficial degree of divergence of all diagrams that is not greater than the order of the bare Lagrangian. On the basis of these results, the renormalization of the theory is achieved by following the procedure developed in Ref.~\cite{Barvinsky:2017zlx}.

We present a summary of the renormalization. Under an inductive approach in the order in loops, it is assumed that at order $(L-1)$ the divergences have been subtracted, such that the action at the $L$th order in loops is given by
\begin{equation}
\Sigma_L = 
\Sigma_{L-1} - \hbar^L \Gamma_{L,\infty} + \mathcal{O}(\hbar^{L+1}) \,,
\label{ansatL}
\end{equation}
where $\Gamma$ is the effective action and $\infty$ stands for its divergent part. In (\ref{ansatL}) and the subsequent analysis, the effective action $\Gamma$ is used as a functional of the quantum fields. 

By following standard procedures, identities on the effective action can be established for this theory. These are the Slavnov-Taylor identity, the Ward identity, and the field equation of the auxiliary field $\pi_i$. In Ref.~\cite{Barvinsky:2017zlx} the general form of these identities, under the background-field approach, can be found. The identities on the effective action imply the following condition of annihilation:
\begin{equation}
\boldsymbol{Q}_+ \hat{\Gamma}_{L,\infty} = 0 \,,
\label{closedgamma}
\end{equation}
where
\begin{eqnarray}
&&
\hat{\Gamma} \equiv
\Gamma 
- \left( \pi_i + \Omega^a \bar{C}_i 
\frac{\delta}{\delta \phi^a} \right)
\left( 
D_t n^i - \rho \Theta^{ijk} h_{jk}
- \rho \mathcal{D}^{ij} ( \pi_j / \sqrt{\bar{g}} ) \right) 
\,,
\\
&&
\boldsymbol{Q}_+ X = 
	\frac{\delta \hat{\Sigma}_0} {\delta \hat{\gamma}_a} 
	\frac{\delta X} {\delta \varphi^a}
	+ \frac{\delta \hat{\Sigma}_0} {\delta \varphi^a} 
	\frac{\delta X} {\delta \hat{\gamma}_a}
	+ \frac{\delta \hat{\Sigma}_0} {\delta J_i} 
	\frac{\delta X} {\delta C^i}
	+ \frac{\delta \hat{\Sigma}_0} {\delta C^i} 
	\frac{\delta X} {\delta J_i}
	+ \Omega^a \frac{\delta X} {\delta \phi^a} \,,
\label{Q+}
\\ &&
\hat{\gamma}_i = 
\mathbb{T}_i - \bar{C}_i D_t \,,
\quad
\hat{\gamma}^{ij} = 
\mathbb{T}^{ij} + \rho \bar{C}_k \Theta^{kij}  \,,
\quad
\hat{\gamma}_a = \gamma_a \; \text{otherwise,} 
\end{eqnarray}
and $\hat{\Sigma}_0$ is the tree-level reduced action
\begin{eqnarray}
 &&
 \hat{\Sigma}_0 = 
 S_0[\varphi^a] + \int dt d^3x\, \boldsymbol{Q} \Psi_{\text{s}} \,,
 \\ &&
 \Psi_{\text{s}} \equiv 
 \tilde{\Psi}
 - \hat{\gamma}_a \varphi^a
 + \bar{J}_i C^i  \,.
\end{eqnarray}
Operator $\boldsymbol{Q}_+$ is nilpotent. The divergence on the effective action $\Gamma_\infty$ is equal to the divergence in $\hat{\Gamma}_\infty$. $\hat{\Sigma}_0$ has no explicit dependence on the background fields.

By using a procedure of expanding $\hat{\Gamma}_{L,\infty}$ on the ghost fields $C^i$ and $\Omega^a$, the cohomology of $\boldsymbol{Q}_+$ and other operators involved in the process guarantee the existence of the functions $\boldsymbol{S}_L[\varphi^a]$ and $\boldsymbol{\Upsilon}_L$, such that the solution of Eq.~(\ref{closedgamma}) at order $L$ in loops is given by
\begin{equation}
\hat{\Gamma}_{L,\infty} = 
\boldsymbol{S}_L[\varphi^a] + \boldsymbol{Q}_+ \boldsymbol{\Upsilon}_L \,.
\end{equation} 
When this solution is substituted in Eq.~(\ref{ansatL}), the form of the $L$th order gauge fermion is identified,
\begin{equation}
\Psi_L = \Psi_{L-1} - \hbar^L \boldsymbol{\Upsilon}_L \,,
\label{psiL}
\end{equation}
as well as the form for the $L$th order counterterm,
\begin{equation}
\Sigma^C_L = 
- \boldsymbol{S}_L[\varphi] 
- \boldsymbol{Q} \boldsymbol{\Upsilon}_L \,.
\end{equation}
The quantum action gets the form
\begin{equation}
\Sigma_L =
S[\varphi] - \sum\limits_{l=1}^L \hbar^l \boldsymbol{S}_l[\varphi] 
+ \boldsymbol{Q} \Psi_L
- \hbar^L \frac{ \delta \boldsymbol{\Upsilon}_L } { \delta \gamma_a }
\frac{ \delta \Sigma_0 } { \delta \varphi^a }
+ \hbar^L \frac{ \delta \boldsymbol{\Upsilon}_L } { \delta \bar{J}_i }
\frac{ \delta \Sigma_0 } { \delta C^i }
+ \mathcal{O}(\hbar^{l+1}) \,.
\end{equation}
The first three terms have the desired manifest BRST structure. In Ref.~\cite{Barvinsky:2017zlx} a field redefinition $\varphi^a,C^i \rightarrow \varphi'^a,C'^i$ is found that eliminates the rest of the terms. It is given by
\begin{equation}
\begin{array}{l}
{\displaystyle
\varphi^a = 
\varphi'^a 
+ \hbar^L \frac{\delta \boldsymbol{\Upsilon}_L } {\delta\gamma_a} (\varphi',C',\ldots) 
+ \mathcal{O}(\hbar^{L+1}) \,, }
\\[2ex] 
{\displaystyle
C^i = 
C'^i 
- \hbar^L \frac{\delta \boldsymbol{\Upsilon}_L } {\delta\bar{J}_i} (\varphi',C',\ldots) 
+ \mathcal{O}(\hbar^{L+1}) \,. }
\end{array}
\label{fieldredefinition}
\end{equation}
After this field redefinition, the $L$th order quantum action acquires the BRST structure
\begin{equation}
\Sigma_L =
S_L[\varphi^a] 
+ \int dt d^3x \, \boldsymbol{Q} \Psi_L \,,
\label{brstrenormalized}
\end{equation}
where $S_L[\varphi^a]$ is a FDiff gauge invariant local functional. The gauge fermion is invariant under background-gauge transformations and has the form
\begin{equation}
\Psi_L =
\bar{C}_i \left( 
D_t n^i - \rho \Theta^{ijk} h_{jk} - \mathcal{D}^{ij} ( \pi_j / \sqrt{\bar{g}} ) \right)
- \gamma_a ( \varphi^a - \phi^a )
+ \bar{J}_i C^i + \mathcal{O}(\hbar^{L+1}) \,.
\end{equation}
In the generating functional $W$, the fields that couple to the external sources, denoted by $\tilde{\varphi}^a$ and $\tilde{C}^i$, 
\begin{equation}
 W_L = 
 -\hbar \log \int \mathcal{D}V \exp\left[ -\frac{1}{\hbar} \left( 
 \Sigma_L + \int dt dt^3 x \left(
 J_a (\tilde{\varphi}^a_L - \phi^a) 
 + \bar{\xi}_i \tilde{C}^i_L 
 + \xi^i \bar{C}_i
 + y^i \pi_i
 \right) \right) \right]  \,,
\end{equation}
are given by the gauge fermion in the form 
\begin{equation}
\tilde{\varphi}^a_L = \phi^a - \frac{\delta \Psi_L}{\delta \gamma_a} \,,
\quad 
\tilde{C}^i_L = \frac{\delta \Psi_L} {\delta\bar{J}_i} \,.
\label{coupledfields}
\end{equation}
The functional relationship (\ref{coupledfields}) is preserved by the field redefinition (\ref{fieldredefinition}) at the $L$th order in loops.

\section{The $(2+1)$-dimensional theory}
In the $(2+1)$-dimensional Ho\v{r}ava theory, the classical action maintains the form (\ref{classicalaction}), adapted to the two spatial dimensions. The $(2+1)$-dimensional case requires a potential of $z=2$ order for power-counting renormalizability. The complete potential with all the inequivalent $z=1,2$ terms compatible with the FDiff symmetry is \cite{Sotiriou:2011dr}
\begin{equation}
\begin{split}
\mathcal{V} =&
-\beta R-\alpha a_i a^i+\alpha_{1}R^{2}+\alpha_{2} (a_i a^i)^2 +\alpha_{3}R a_i a^i +\alpha_{4} a_i a^i \nabla_k a^k
\\ &
+\alpha_{5} R\nabla_{i}a^{i} 
+\alpha_{6}\nabla^{i}a^{j}\nabla_{i}a_{j}+ \alpha_{7}(\nabla_{i}a^{i})^{2} \,.
\end{split}
\end{equation}
The coupling constants of the theory are $\lambda$, $\beta$, $\alpha$, and $\alpha_1$,...,$\alpha_7$. We hope that the repetition of the notation on various constants in the $(3+1)$ and $(2+1)$ cases does not cause confusion. The primary classical Hamiltonian has the same functional form (\ref{H0}), with the combinations of constants: 
\begin{equation}
\sigma = \frac{1-\lambda}{1-2\lambda} \,,
\quad
\bar{\sigma} = \frac{\lambda}{1-2\lambda} \,,
\end{equation}
and we also use $\alpha_{67} = \alpha_6 + \alpha_7$. The two second-class constraints are the vanishing of the momentum conjugate to $N$, and the constraint
\begin{equation}
\begin{split}
\theta_{1} \equiv\, &  
\sqrt{g} N \left( \frac{\pi^{ij}\pi_{ij}}{g} 
+ \bar{\sigma} \frac{\pi^{2}}{g} + \mathcal{V} \right) 
+ \sqrt{g} \Big( 
2\alpha\nabla_{i}(Na^{i})
- 4\alpha_{2} \nabla_{i} ( N a^{i} a_k a^k )
- 2\alpha_{3}\nabla_{i}(NRa^{i})
\\&
+ \alpha_{4} \left( \nabla^{2}( N a_k a^k ) 
-2\nabla_{i}( N a^{i} \nabla_{j} a^{j} ) \right)
+ \alpha_{5}\nabla^{2}(NR)
+ 2\alpha_{6}\nabla^{i}\nabla^{j}(N\nabla_{j}a_{i})
\\ &
+ 2\alpha_{7}\nabla^{2}(N\nabla_{i}a^{i})
\Big) 
= 0 \,.
\end{split}
\end{equation}
Similar to the $(3+1)$ case, the primary Hamiltonian (\ref{H0}) is equivalent to the integral of this second-class constraint, $H_{0} = \int d^{2}x \,\theta_{1}$.

The BFV quantization \cite{Bellorin:2022qeu} is done by repeating the same steps of the $(3+1)$ case. In particular, the measure of the second-class constraints maintain the same functional structure (\ref{measureA}), as well as the BRST charge $\Omega$ and the BFV gauge fermion $\Psi$. The factor $\chi^i$ is defined by
\begin{equation}
\chi^{i}= 
\mathfrak{D}^{ij}\pi_{j} 
- 2 \Delta\partial_jh_{ij}
+ 2 \lambda \bar{\kappa} \Delta\partial_{i}h
- 2\kappa \partial_{ijk} h_{jk}
\,,
\label{chi}
\end{equation}
where $\mathfrak{D}^{ij} = \delta_{ij} \Delta + \kappa \partial_{ij}$. By performing the integration on the BFV ghosts $\mathcal{P}^i,\bar{\mathcal{P}}_i$ and (re)defining the BRST transformations in the same way, the quantum action can be written in the BRST-invariant form (\ref{quantumactionbrst}).

The propagators of the resulting $(2+1)$ quantum theory are
\begin{equation}
\begin{split}
&	\langle \bar{C}_iC^{j}\rangle = - \mathcal{S}_{ij}  \,,
	\\
&	\langle p^{ij}p^{kl} \rangle =
	\rho_2 k^{4} P_{ij} P_{kl}\mathcal{T}_{1} \,,
	\\
&	\langle h_{ij} h_{kl} \rangle =
	8 \left(\omega^2 - 2\rho k^{4}\right) Q_{ijkl}\mathcal{T}^{2}_{3}
	+ 4 \left( {\sigma} P_{ij} P_{kl}
	+ \bar{\sigma} \hat{k}_i \hat{k}_j P_{kl} 
	+ \bar{\sigma} \hat{k}_k \hat{k}_l P_{ij} 
	\right) \mathcal{T}_{1}
	+ 4 \sigma Q \hat{k}_{i} \hat{k}_{j} \hat{k}_{k} \hat{k}_{l} \mathcal{T}_{1} \mathcal{T}_{2}^{2} \,,
	\\
&	\langle h_{ij} p^{kl} \rangle =
	2\omega \left[ 2Q_{ijkl} \mathcal{T}_{3}
	+ P_{ij} P_{kl}\mathcal{T}_{1}
	+ 4 \rho \bar{\sigma}\bar{\kappa} \left(1- 2\lambda +  \rho_2 \right) k^2 k_{i}k_{j} P_{kl} \mathcal{T}_{1}\mathcal{T}_{2}
	+  \hat{k}_{i} \hat{k}_{j} \hat{k}_{k} \hat{k}_{l} \mathcal{T}_{2} \right] \,,
	\\
&	\langle n^{i} h_{jk} \rangle =
	16 i \omega \rho k^3 \left( P_{i(j} \hat{k}_{k)} \mathcal{T}_{3}^{2}
	+ \bar{\kappa} \hat{k}_i \hat{k}_j \hat{k}_k \mathcal{T}_{2}^{2} \right) \,,
	\\
&	\langle n^{i} p^{jk} \rangle =
	2 i k^3 \left[ 2 P_{i(j} \hat{k}_{k)} \mathcal{T}_{3}
	+ \hat{k}_{i} \left( \rho_1 \hat{k}_j \hat{k}_{k}
	- 2 \lambda\bar{\kappa} P_{jk} \right) \mathcal{T}_{2} 
	\right] \,,
	\\
&	\langle n^{i} n^{j} \rangle =
	- 4 \rho k^2 P_{ij} \left(\omega^{2}-2\rho k^4\right)\mathcal{T}^{2}_{3}
	- 4 \rho \bar{\kappa} \left(\omega^{2} - 2\rho\rho_{1}k^{4} \right) k_{i}k_{j} \mathcal{T}_{2}^{2} \,,
	\\ 
&	\langle h_{ij} n \rangle =
	\frac{2\alpha_{5}}{\alpha_{67}} \left( \sigma P_{ij}
	+ \bar{\sigma} \hat{k}_i \hat{k}_j \right)\mathcal{T}_{1} \,,
	\quad
	\langle p^{ij} n \rangle =
	\frac{\alpha_{5}}{\alpha_{67}}\omega P_{ij}\mathcal{T}_{1} \,,
	\quad
	\langle n^{i}\pi_{j} \rangle =
	\omega \mathcal{S}_{ij}  \,,
	\\
&	\langle \pi_i h_{jk} \rangle =
	- 4 i \left( P_{i(j} k_{k)} \mathcal{T}_{3}
	+ k_{i} \hat{k}_{j} \hat{k}_{k}\mathcal{T}_{2} \right) \,,
	\quad
	\langle nn \rangle =
	\frac{ \alpha_{5}^2 \sigma }{ \alpha_{67}^2 } \mathcal{T}_{1} \,,
	\end{split}
	\label{propagators2d}
\end{equation}
and 
\begin{equation}
	\langle \mathcal{A}\mathcal{A} \rangle =
	\langle \mathcal{A} n \rangle = 
	\langle \bar{\eta} \eta \rangle = 
	\frac{1}{ \alpha_{67} k^{4}} \,,
\end{equation}
where
\begin{equation}
	\begin{split}
	&
	\hat{k}_i = \frac{ k_i }{k} \,,
	\quad
	P_{ij} = \delta_{ij} - \hat{k}_{i} \hat{k}_{j} \,,
	\\ &
	Q_{ijkl} =
	\hat{k}_i \hat{k}_k P_{jl} + \hat{k}_j \hat{k}_k P_{il} 
	+ \hat{k}_i \hat{k}_l P_{jk} + \hat{k}_j \hat{k}_l P_{ik} \,,
	\\ &
	Q =
	\omega^{4} 
	+ \left[ 4\rho\left(2\lambda^2 +2\lambda -1 \right)\bar{\kappa}
	- \rho_2 \right] 
	\frac{\omega^{2}k^{4}}{1-\lambda} 
	+ 4 \bar{\kappa} \rho \left( \rho_2 + 4\lambda^{2}\bar{\kappa}\rho \right) k^8  \,,
	\\ &
	\mathcal{T}_{1} =
	\left(\omega^2-\rho_2 {\sigma}k^4\right)^{-1} \,,
	\quad
	\mathcal{T}_{2} = 
	\left(\omega^{2} + 2 \rho\rho_1 k^4 \right)^{-1},
	\\ &
	\mathcal{T}_{3} =
	\left(\omega^{2} + 2\rho k^{4}\right)^{-1},
	\quad
	\mathcal{S}_{ij} =
	2 P_{ij}\mathcal{T}_{3} + 2 \hat{k}_{i} \hat{k}_{j} \mathcal{T}_{2} \,,
	\\ &
	\rho_{1} = 2( 1 - \lambda )( 1 + \kappa ) \,,
	\quad
	\rho_{2} = 4\alpha_{1} - \frac{ \alpha^{2}_{5} }{ \alpha_{67} } \,.
	\end{split}
\end{equation}
All the propagators in the list (\ref{propagators2d}) satisfy the condition of regularity, whenever the constants satisfy the following bounds:
\begin{equation}
	\rho > 0\,, 
	\quad 
	( 1 - \lambda ) ( 1 + \kappa ) > 0 \,, 
	\quad 
	\left( \frac{1-\lambda}{1-2\lambda} \right) 
	\left( 4 \alpha_1 - \frac{\alpha_5^2} { \alpha_6 + \alpha_7 } 
	\right) < 0 \,.
\end{equation}
As in the $(3+1)$ case, the propagators involving the fields $\mathcal{A},\eta,\bar{\eta}$ are irregular. The degree of superficial divergence of this theory is $D_{\text{div}} \leq 4$. Cancellation of irregular loops works in the same fashion as the $(3+1)$ case \cite{Bellorin:2022qeu,Bellorin:2023dwk}. A similar background-field formulation can be done. With these results, the renormalization of the $(2+1)$-dimensional Ho\v{r}ava theory is achieved by following the same steps.

\section{Conclusions}
We have presented the proof of renormalization of the nonprojectable Ho\v{r}ava theory, adopting the background-field approach developed by Barvinsky et al.~for gauge theories \cite{Barvinsky:2017zlx}. This approach implies that, once the counterterms have been considered, the renormalized action preserves the BRST structure with the background fields.

We have studied the theory in 3 spatial dimensions, managing all the inequivalent terms of the Lagrangian that are dominant at the ultraviolet (the $z=3$ terms), and contribute to the propagators. We highlight the fact that we have arrived at a manifest BRST invariant form of the action within the Hamiltonian formalism of the BFV quantization; only some integrations on ghost canonical momenta were required. This holds in part thanks to the compatibility of the Hamiltonian formalism with the FDiff symmetry, as indeed happens in the classical theory (and even in general relativity). Moreover, we have succeeded in implementing the BRST transformations of the variables associated with the measure of the second-class constraints, in such a way that all these variables enter in the BRST symmetry on the same footing as the rest of the variables. Another crucial ingredient for the proof is the previous result about the cancellation of irregular loops. The multiplicative divergences on the direction of the frequency produced by irregular loops are the only dangerous effects of the irregular propagators. But they cancel themselves completely. In the direction of the spatial momentum, these propagators behave in the same way as the regular ones. 

The nonprojectable Ho\v{r}ava theory is a consistent quantum theory, unitary and renormalizable, that might yield the classical dynamics of general relativity at the limit of large distances, where higher-order derivatives can be neglected. Once the proof of renormalization has been given, studies on the renormalization flow can be undertaken. A natural question is whether the flow of the coupling constants of the $z=1$ terms, which are the ones dominant at large distances, tends to the case when the theory reproduces general relativity.



\begin{thebibliography}{99}
	\bibitem{Horava:2009uw} 
	P.~Ho\v{r}ava,
	Quantum Gravity at a Lifshitz Point,
	Phys.\ Rev.\ D {\bf 79} 084008 (2009) 
	[arXiv:0901.3775 [hep-th]].
	
	\bibitem{Blas:2009qj} 
	D.~Blas, O.~Pujolas and S.~Sibiryakov,
	Consistent Extension of Ho\v rava Gravity,
	Phys.\ Rev.\ Lett.\  {\bf 104} 181302 (2010)
	[arXiv:0909.3525 [hep-th]].
	
	\bibitem{Bellorin:2022efu}
	J.~Bellor\'in, C.~B\'orquez and B.~Droguett,
	BRST symmetry and unitarity of the Ho\v{r}ava theory,
	Phys. Rev. D \textbf{107} 044059 (2023)
	[arXiv:2212.14079 [hep-th]].
	
	\bibitem{Fradkin:1975cq}
	E.~S.~Fradkin and G.~A.~Vilkovisky,
	Quantization of relativistic systems with constraints,
	Phys. Lett. B \textbf{55} 224 (1975).
	
	\bibitem{Batalin:1977pb}
	I.~A.~Batalin and G.~A.~Vilkovisky,
	Relativistic S Matrix of Dynamical Systems with Boson and Fermion Constraints,
	Phys. Lett. B \textbf{69} 309 (1977).
	
	\bibitem{Fradkin:1977xi}
	E.~S.~Fradkin and T.~E.~Fradkina,
	Quantization of Relativistic Systems with Boson and Fermion First and Second Class Constraints,
	Phys. Lett. B \textbf{72} 343 (1978).
	
	\bibitem{Barvinsky:2015kil}
	A.~O.~Barvinsky, D.~Blas, M.~Herrero-Valea, S.~M.~Sibiryakov and C.~F.~Steinwachs,
	Renormalization of Ho\v{r}ava gravity,
	Phys. Rev. D \textbf{93} 064022 (2016)
	[arXiv:1512.02250 [hep-th]].
	
	\bibitem{Bellorin:2021udn}
	J.~Bellor\'in, C.~B\'orquez and B.~Droguett,
	Quantum Lagrangian of the Ho\v{r}ava theory and its nonlocalities,
	Phys. Rev. D \textbf{105} 024065 (2022)
	[arXiv:2112.10295 [hep-th]].
	
	\bibitem{Senjanovic:1976br}
	P.~Senjanovic,
	Path Integral Quantization of Field Theories with Second Class Constraints,
	Annals Phys. \textbf{100} 227 (1976)
	[erratum: Annals Phys. \textbf{209} 248 (1991)].
	
	\bibitem{Fradkin1973}
	E.~S.~Fradkin,
	Acta Universitatis Wratislaviensis No. 207, in Proceedings of X-th Winter School of Theoretical Physics in Karpacz (Wydawnictwo Uniwersytetu Wroclawskiego Sp., Poland, 1973).
	
	\bibitem{Bellorin:2022qeu}
	J.~Bellor\'in, C.~B\'orquez and B.~Droguett,
	Cancellation of divergences in the nonprojectable Ho\v{r}ava theory,
	Phys. Rev. D \textbf{106} 044055 (2022)
	[arXiv:2207.08938 [hep-th]].
	
	\bibitem{Bellorin:2023dwk}
	J.~Bellor\'in, C.~B\'orquez and B.~Droguett,
	Effective action of the Ho\v{r}ava theory: Cancellation of divergences,
	Phys. Rev. D \textbf{109} 084007 (2024)
	[arXiv:2312.16327 [hep-th]]. 
	
	\bibitem{Anselmi:2007ri}
	D.~Anselmi and M.~Halat,
	Renormalization of Lorentz violating theories,
	Phys. Rev. D \textbf{76} 125011 (2007)
	[arXiv:0707.2480 [hep-th]].
	
	\bibitem{Anselmi:2008bq}
	D.~Anselmi,
	Weighted power counting and Lorentz violating gauge theories. I. General properties,
	Annals Phys. \textbf{324} 874 (2009)
	[arXiv:0808.3470 [hep-th]].
	
	\bibitem{Barvinsky:2017zlx}
	A.~O.~Barvinsky, D.~Blas, M.~Herrero-Valea, S.~M.~Sibiryakov and C.~F.~Steinwachs,
	Renormalization of gauge theories in the background-field approach,
	JHEP \textbf{07} 035 (2018)
	[arXiv:1705.03480 [hep-th]].
	
	\bibitem{Bellorin:2012di}
	J.~Bellor\'in, A.~Restuccia and A.~Sotomayor,
	Non-perturbative analysis of the constraints and the positivity of the energy of the complete Horava theory,
	Phys. Rev. D \textbf{85} 124060 (2012)
	[arXiv:1205.2284 [hep-th]].
	
	\bibitem{DeWitt:1967ub}
	B.~S.~DeWitt,
	Quantum Theory of Gravity. 2. The Manifestly Covariant Theory,
	Phys. Rev. \textbf{162} 1195 (1967); Quantum Theory of Gravity. 3. Applications of the Covariant Theory, Phys. Rev. \textbf{162} 1239 (1967).
	
	\bibitem{Abbott:1981ke}
	L.~F.~Abbott,
	Introduction to the Background Field Method,
	Acta Phys. Polon. B \textbf{13} 33 (1982).
	
	\bibitem{Kluson:2010nf}
	J.~Kluson,
	Note About Hamiltonian Formalism of Healthy Extended Ho\v{r}ava-Lifshitz Gravity,
	JHEP \textbf{07} 038 (2010)
	[arXiv:1004.3428 [hep-th]].
	
	\bibitem{Donnelly:2011df}
	W.~Donnelly and T.~Jacobson,
	Hamiltonian structure of Ho\v{r}ava gravity,
	Phys. Rev. D \textbf{84} 104019 (2011)
	[arXiv:1106.2131 [hep-th]].
	
	\bibitem{Bellorin:2011ff}
	J.~Bellor\'in and A.~Restuccia,
	Consistency of the Hamiltonian formulation of the lowest-order effective action of the complete Ho\v{r}ava theory,
	Phys. Rev. D \textbf{84} 104037 (2011)
	[arXiv:1106.5766 [hep-th]].
	
	\bibitem{Colombo:2014lta}
	M.~Colombo, A.~E.~Gumrukcuoglu and T.~P.~Sotiriou,
	Ho\v{r}ava gravity with mixed derivative terms,
	Phys. Rev. D \textbf{91} 044021 (2015)
	[arXiv:1410.6360 [hep-th]].
	
	\bibitem{Bellorin:2021tkk}
	J.~Bellor\'in and B.~Droguett,
	BFV quantization of the nonprojectable (2+1)-dimensional Ho\v{r}ava theory,
	Phys. Rev. D \textbf{103} 064039 (2021)
	[arXiv:2102.04595 [hep-th]].
	
	\bibitem{Sotiriou:2011dr}
	T.~P.~Sotiriou, M.~Visser and S.~Weinfurtner,
	Lower-dimensional Horava-Lifshitz gravity,
	Phys. Rev. D \textbf{83} 124021 (2011)
	[arXiv:1103.3013 [hep-th]].
	
\end{thebibliography}
\end{document}